\documentclass[fleqn,10pt]{wlpeerj}

\usepackage{graphicx}
\usepackage{latexsym}
\usepackage{framed}

\usepackage{color}
\usepackage{url}
\usepackage{colortbl}

\definecolor{darkblue}{rgb}{0.0,0,0.5} 
\newcommand{\kw}[1]{{\tt {\color{darkblue} #1}}}

\title{Implementing a Small Parsing Virtual Machine on Embedded Systems}

\author[1]{Shun Honda}
\author[1]{Kimio Kuramitsu}
\affil[1]{GraduateSchool of Electronic and Computer Engineering, Yokohama National University}

\keywords{Schema validation, XML/DTD, parsing expression grammars, and performance}

\begin{abstract}
PEGs are a formal grammar foundation for describing syntax, and are not hard to generate parsers with a plain recursive decent parsing. However, the large amount of C-stack consumption in the recursive parsing is not acceptable especially in resource-restricted embedded systems. Alternatively, we have attempted the machine virtualization approach to PEG-based parsing. MiniNez, our implemented virtual machine, is presented in this paper with several downsizing techniques, including instruction specialization, inline expansion and static flow analysis. As a result, the MiniNez machine achieves both a very small footprint and competitive performance to generated C parsers. We have demonstrated the experimental results by comparing on two major embedded platforms: Cortex-A7 and Intel Atom processor.
\end{abstract}

\begin{document}

\flushbottom
\maketitle
\thispagestyle{empty}

\section{Introduction}

In recent years, the increased use of text data on the Web and IoTs raises new demands for processing texts on embedded systems. Text data are flexible in data representation, and can carry a variety of data formats such as XML and sensor logs. To  make the use of such various formats, we need advanced text processing, that is, {\em parsing} and {\em syntactic analysis}. 

In developing embedded systems, C is a popular programming language. In parallel, it is a known fact that C suffers from poor language supports for strings. Unfortunately, the poorness has enforced C developers to heavily use low-level pointer operations, resulting in a major source of bugs and security vulnerability. Regular expression libraries, such as PCRE library\cite{Pcre}, have reduced such error-prone coding, but still suffers from its limited expressiveness for handling structured data. 

We address the use of Parsing Expression Grammars\cite{POPL04_PEG} on resource-restricted embedded systems. PEGs are a new recognition-based formal grammar and more expressive than regular expressions, so that the use of PEGs would enable us to parse and extract various text data, including HTML, XML, JSON and various logs, without any program code. 

The main challenge of this paper is that we attempt to an small implementation of virtual parsing machine. The reason is that the recursive decent parsing, a common algorithm of PEG-based parsing, requires the uncontrolled size of the C stack consumption for recursion. A large amount of C stack consumption is not definitively suitable for embedded systems since the maximum stack size is limited to 4K-16K bytes. 
To avoid uncontrolled C stack consumption, the support of a controllable specialized stack is attempted with a machine virtualization. 

MiniNez is a small virtual machine designed for PEG parsing. The MiniNez works with optimized bytecode that is produced by a MiniNez compiler. Due to the separation of compiler and code optimizer, the resulting MiniVM machine achieves a very small footprint, implemented with approximately 100 lines in C source code. Despite such compactness, MiniNez provides us with rich parsing capability for various type of text data, including CSV, XML, JSON, Syslog, and others. In addition, MiniNez achieves a competitive performance, compared to generated C parsers. 

The remainder of the paper is structured as follows. Section 2 introduce PEGs and opportunities for text data. Section 3 presents the design of stackless virtual machine. Section 4 describes the detailed implementation of MiniNez machine with several compilation techniques for reduced code size.
Section 5 demonstrates the experimental results. 
Section 6 reviews related work. 
Section 7 concludes the paper. 

\section{Parsing Expression Grammars}

Parsing Expression Grammars (PEGs) are recognition-based foundation for describing syntax, formalized by Ford\cite{POPL04_PEG}. 

\subsection{Notion and Operators}

A PEG is a collection of production rules, which take an EBNF-like form $A = e$, where $A$ is a nonterminal and $e$ is a parsing expression. Parsing expressions are composed by PEG operators, which looks like regular expressions.

\begin{table}[bt]
\begin{center}
\caption{PEG Operators} \label{table:PEGs}
\begin{tabular}{llll} \hline
PEG  & Type & Proc. & Description\\ \hline
\verb|' '| & Primary & 5 & Matches text\\
$[ ]$ & Primary & 5 & Matches character class \\
$.$ & Primary & 5 & Any character\\
$A$ & Primary & 5 & Non-terminal application\\
$( e )$ & Primary & 5 & Grouping\\
$e?$ & Unary suffix & 4 & Option\\
$e*$ & Unary suffix & 4 & Zero-or-more repetitions\\
$e+$ & Unary suffix & 4 & One-or-more repetitions\\
$\&e$ & Unary prefix & 3 & And-predicate\\
$!e$ & Unary prefix & 3 & Negation\\
$e_1 e_2$ & Binary & 2 & Sequencing\\
$e_1 / e_2$ & Binary & 1 & Prioritized Choice\\ \hline
\end{tabular}
\end{center}
\end{table}

Table \ref{table:PEGs} shows a summary of PEG operators that compose an expression. The string 'abc' exactly matches the same input, while [abc] matches one of these characters. The . operator matches any single character. The lexical match consumes the matched size of characters and moves forward a position of matching. The $e?$, $e*$, and $e+$ expressions behave as in common regular expressions, except that they are greedy and matches until the longest position. The $e_1\;e_2$ attempts two expressions $e_1$ and $e_2$ sequentially, backtracking the starting position if either expression fails.  The choice $e_1\;/\; e_2$ first attempt $e_1$ and then attempt $e_2$ if $e_1$ fails. The expression $\&e$ attempts $e$ without any character consuming. The expression $!e$ fails if $e$ succeeds, but fails if $e$ succeeds. Furthermore information on PEG notations are referred as to \cite{POPL04_PEG}.

\subsection{Opportunities for Text Data} 

PEGs have been originally formalized for programming language parsers. However, the formalism of PEGs is composed by several desirable properties for parsing text data. Here we introduce the properties of PEGs with example several grammars for text data. 

Figure \ref{fig:csv} shows a PEG grammar for the syntax of Comma Separated Value, or simply CSV, where values are simply parsed by the separation character such as \verb|','| and \verb|'\n'|. Although this grammar is fairly simplified for readability, this suggests how small set of productions make possible a lot. Note that PEGs are a scanner-less parsing, so that both lexical and syntax analysis are specified in an integrated way. 

\begin{figure}[tb]
\begin{small}
\begin{tabular}{p{1.5cm} l}
{\tt File} & \verb|= CSV*| \\
{\tt CSV} & \verb|= Value ( ',' Value )* '\n'| \\
{\tt Value} & \verb|= (![,\n] .)*| \\
\end{tabular}
\end{small}
\caption{Syntax Definition of CSV}
\label{fig:csv}
\end{figure}

\begin{figure}[tb]
\begin{small}
\begin{tabular}{p{1.5cm} l}
{\tt TopLevel} & \verb|= PROLOG? _* DTD? _* Element _*| \\ 
{\tt PROLOG} & \verb|=  '<?xml' (!'?>' .)* '?>'| \\ 
{\tt DTD} & \verb|= '<!' (!'>' .)* '>'| \\
{\tt Element  } & \verb|= '<' Name (_+ Attribute)* ('/>' / '>'| \\ 
            & \verb|    Content '</' Name '>') _*| \\
{\tt Name} & \verb|= [A-Za-z:] ('-' / [A-Za-z0-9:._])*| \\
{\tt Attribute} & \verb|= Name _* '=' _* String| \\
{\tt String} & \verb|= '"' (!'"' .)*  '"'| \\
{\tt Content } & \verb|= (Element / CDataSec / CharData)*| \\ 
{\tt CDataSec  } & \verb|=  '<![CDATA['  (!']]>' .)* ']]>' _*| \\
{\tt COMMENT  } & \verb|= '<!--' (!'-->' .)* '-->' _*| \\ 
{\tt CharData  } & \verb|=  (!'<' .)+| \\ 
\verb|_| & \verb|= [ \t\r\n]| \\ 
\end{tabular}
\end{small}
\caption{Syntax Definition of XML}
\label{fig:xml}
\end{figure}

Figure \ref{fig:xml} shows a PEG grammar for the XML 1.0 syntax. Due to EBNF-like nonterminal calls, PEG can recognize recursively nested structures, which is hardly recognized by regular expressions. In addition, PEG's unlimited look-aheads by syntactic predicates (such as $\&e$ and $!e$) allow recognizing such a nested structure as $\{a^n b^n c^n | n > 0\}$, which cannot be recognized in CFGs. 

Figure \ref{fig:syslog} is a grammar fragment for the syslog format. PEGs can describe the content of data elements such as {\tt DATE} and {\tt TIME} with lexical patterns, as well as the syntactic structure. This feature helps analyze unstructured text data. In addition, it is allowed to compose defined nonterminals from other PEGs, since PEGs are closed under composition (notably, intersection and complement). This means we can easily compose two different grammars to parse a mixed content such as log entries containing an XML element. 

\begin{figure}[tb]
\begin{small}
\begin{tabular}{p{1.5cm} l}
{\tt File} & \verb|= Log*| \\ 
{\tt Log  } & \verb|= MONTH ' '  DAY ' ' TIME ' ' HOST ' ' | \\  
     & \verb|  PROCESS '[' PID ']' Misc ': ' DATA| \\     
{\tt DAY } & \verb|= [0-3 ][0-9]| \\ 
{\tt MONTH  } & \verb|= 'Jan'/'Feb'/'Mar'/'Apr'/'May'/'Jun'| \\ 
      & \verb|  /'Jul'/'Aug'/'Sep'/'Oct'/'Nov'/'Dec'| \\ 
{\tt TIME   } & \verb|= [0-9][0-9] ':' [0-9][0-9] ':' [0-9][0-9]| \\ 
{\tt HOST } & \verb|= (!' ' .)*| \\ 
{\tt PROCESS  } & \verb|= (!'[' .)*| \\ 
{\tt PID } & \verb|= [0-9]+| \\ 
{\tt DATA } & \verb|= (!('\n' (MONTH / !.)) .)*| \\ 
\end{tabular}
\end{small}
\caption{Grammar of Syslog Format}
\label{fig:syslog}
\end{figure}

\begin{figure}[tb]
\begin{small}
\begin{tabular}{p{1.5cm} l}
{\tt File  } & \verb| = ((!EMAIL .)* EMAIL )*| \\
{\tt EMAIL  } & \verb| = LOCAL '@' DOMAIN| \\
{\tt LOCAL  } & \verb| = ([A-Za-z0-9] / '-')+ ('.' LOCAL)?| \\
{\tt DOMAIN  } & \verb| = SUBDOMAIN ('.' SUBDOMAIN)+| \\
{\tt SUBDOMAIN  } & \verb| = ([A-Za-z0-9] / '-')+| \\
\end{tabular}
\end{small}
\caption{Example of Information Extraction}
\label{fig:email}
\end{figure}

The last example comes from an application to the information extraction. Figure \ref{fig:email} is a PEG for finding all emails from a text. Unlike the traditional parser construction defining a complete set of syntax rules, we can treat an interesting part of text data. Note that PEGs provide the recognition capability only. That is, another complemental trick is needed to extract from the results of the recognition. Several approaches \cite{PLDI06_Rats,LPeg,ASTMachine} have been proposed and implemented in contexts of PEG-based tools. 

\subsection{Problems}

PEGs have been mostly used in a parser generation scenario. That is, the grammar specification is converted to a parser  source code written in C or other general-purpose programming language. Since PEGs can be simply implemented with a recursive decent parsing, the parser generation is easier than that in other formal grammars such as LALR, and fits in compaction requirements for embedded systems.  

To illustrate the parser generation, let us consider the following simple production rule:  

{\small \begin{verbatim}

  A = 'a' A / ''

\end{verbatim} }

The production A is converted into a nonterminal function which looks like:

{\small \begin{verbatim}

  bool A() {
    if(match('a')) {
      return A();
    }
    return true;
  }

\end{verbatim} }

A problem with a recursive decent parsing is that every nonterminal call requires the C stack consumption. Let us take a look back at the above example; the necessary amount of C stack depends on the length of a given input {\tt 'a'} character sequence. That is, we need 5 recursive nonterminal calls of A to match {\tt 'aaaaa'}. In reality, the grammar developer mostly uses non-recursive expression {\tt 'a'*} for repetition, the total amount of stack consumption can fit in a standard C stack size. However, in embedded settings that are characterized  in common by very limited C stack space (such as 4KB), the recursive decent parser would easily consume all available C stack during parsing. 

\section{Parsing Virtual Machine}

The parsing virtual machine is an alternative implementation that makes no use of C stack. Nonterminal calls are replaced with VM instructions that operate over independently-allocated VM stacks. The consumption of VM stacks can be smaller than that of C stacks, since we can specialize VM stack operations based on PEG parsing. Moreover, the user can control the size of VM stack allocation in order to fit the grammar specification.

In this section we describe our motivation of {\em stackless} virtual machine, a set of designed VM instructions, and the conversion from parsing expressions to designed instructions. Note that the term "stackless" in this paper implies that no C stacks are consumed when parsing. 

\subsection{Parsing Instructions}

Parsing instruction is an abstract VM instruction that is selected to compose a minimum instruction set to execute all PEGs operators. Our designed instruction set consists of 12 instructions, summarized in Table \ref{table:core}. 

In the past, LPeg\cite{DLS08_LPeg}, another parsing machine implementation, defines a similar set of parsing instructions. However, the design principle differ from ours in terms of backtracking handling. LPeg's instruction set has adopted {\em dynamic global indirect jump} to alternative. As in an exception handling, a jump address to alternative code is pushed on the stack and looked up at backtracking time. On the contrary, our instruction set is based on {\em static local direct jump} to alternative. The jump address is determined at the compilation time. As a result, we expect that the amount of stack consumption can be further reduced. 

\begin{table}[tbh]
\caption{Instruction set in parsing machine}
\label{table:core}
\begin{center}
\begin{tabular}{lll}
Opcode & Argument & Description \\ \hline
\kw{nop} & &
Do nothing \\
\kw{succ} & &
Succed always \\
\kw{fail} & &
Fail always \\
\kw{char} & c &
Match the character c \\
\kw{any} &  &
Match any character \\
\kw{jump} & L &
Branch always \\
\kw{iffail} & L &
Branch if the result is {\tt F} \\
\kw{call} & L &
Call the nonterminal call \\
\kw{ret} &  &
Return the nonterminal call \\
\kw{push} &  &
Push $pos$ to the stack \\
\kw{pop} &  &
Pop the top operand stack value \\
\kw{peek} &  &
Store the top operand stack value \\ 
\hline
\end{tabular} \\
L: Label representing a position of code point
\end{center}
\end{table}

\begin{figure}[tbh]

\[
\frac{~~~~ \kw{nop}() ~~~~ (pc, pos, sp, r) ~~~~}
     {(pc+1, pos, sp, r) } 
\]

\[
\frac{~~\kw{char}(c) ~~ (pc, pos, sp, \top) ~~ I[pos]=c ~~}
     {(pc+1, pos+1, sp, \top) } , ~
\frac{~ \kw{char}(c) ~~ (pc, pos, sp, \top) ~~ I[pos] \neq c ~~}
     {(pc+1, pos, sp, \bullet) } 
\]

\[
\frac{~~~ \kw{call}(L) ~~~~ (pc, pos, sp, \top) ~~~ }
     {~~~~ (L, pos, stack.push(pc+1), \top) ~~~~ }
     ~~~~
\frac{~~~ \kw{ret}() ~~~~ (pc, pos, sp, r) ~~~}
     {(stack[sp], pos, sp-1, r) } 
\]

\[
\frac{~~~ \kw{iffail}(L) ~~~~ (pc, pos, sp, \top) ~~~ }
     {~~~~ (pc+1, pos, sp, \top) ~~~~ }, ~~
\frac{~~~ \kw{iffail}(L) ~~~~ (pc, pos, sp, \bullet) ~~~}
     {(L, pos, sp, \bullet) } 
\]

\[
\frac{~~~ \kw{succ}() ~~~~ (pc, pos, sp, r) ~~~ }
     {~~~~ (pc+1, pos, sp, \top) ~~~~ }, ~~ 
\frac{~~~ \kw{fail}() ~~~~ (pc, pos, sp, r) ~~~}
     {(pc+1, pos, sp, \bullet) } 
\]

\[
\frac{~~~ \kw{push}() ~~~~ (pc, pos, sp, r) ~~~ }
     {~~~~ (pc+1, pos, stack.push(pos), r) ~~~~ } ~~~ 
\frac{~~~ \kw{pop}() ~~~~ (pc, pos, sp, r) ~~~ }
     {~~~~ (pc+1, pos, sp-1, r) ~~~~ } 
\]
\[
\frac{~~~ \kw{peek}() ~~~~ (pc, pos, sp, r) ~~~}
     {(pc+1, stack[sp], sp, r) } 
\]

\caption{Operational Semantics of Parsing Instructions}
\label{fig:semantics}
\end{figure}

The semantics of parsing instructions can be defined as a state transition of a parser instance. The instance has four states, denoted $pos, pc, sp, r$, which are defined respectively:

\begin{itemize}
\item $pos$ -- a position at which the parser matches the input stream 
\item $pc$ -- program counter
\item $sp$ -- an index of the operand $stack$ top 
\item $r$ -- a Boolean value for representing either $\top$ (success) or $\bullet$ (failure). 
\end{itemize}

Let $I$ be a input characters and $I[pos]$ be a character at $pos$-th position. We use $L, L', L''$ to range over labels for a jump pointer. Figure \ref{fig:semantics} shows the operational semantics of parsing instructions.

Backtracking is a notorious behavior in PEG-based parsing. The parsing machine must move back to a parsing position to execute alternative instructions. In our instruction set, we save a current parsing position $pos$ by \kw{push}, and restore it by \kw{peek}. The \kw{pop} is used to rollback the stack status. All parsing expressions that need to handle backtracking are surrounded by \kw{push} and \kw{pop}.  In backtracking cases, we restore the pushed position, and use \kw{succ}() to turn the parsing result into success. Figure \ref{fig:succ} depicts a basic flow of success and failure case of parsing. 

\begin{figure}[tbh]
{\large
	\begin{verbatim}	
            [SUCCESS]     [FAIL]          
              push()      
              ...
              pop()       peek()
                          pop()
                          succ()
	\end{verbatim}
}
\caption{A Basic Flow of Failure Handling}
\label{fig:succ}
\end{figure}

\subsection{Converting Parsing Expressions}

The conversion from parsing expressions to parsing instructions can be defined in an inductive manner. Let $\tau(e, L)$ be a conversion function that maps from a given parsing expression $e$ to an instruction sequence. The second argument $L$ is a jump address that is performed in the failure evaluation case of $e$. Figure \ref{fig:nez} is a definition of an abstract syntax of parsing expression $e$. Figure \ref{fig:convert} shows a full conversion of all parsing expressions to sequence of parsing instructions. 

\begin{figure}[tbh]
\begin{center}

\begin{tabular}{|lrrl|} \hline
$\tau(\epsilon,\: L)$      &  \verb|      = | & & $\kw{nop}$  \\
$\tau(c,\: L)$      &  \verb|      = | & & $\kw{char} ~c$  \\
 &  & & $\kw{iffail}(L)$ \\
 $\tau(.,\: L)$      &  \verb|      = | & & $\kw{any}$  \\
 &  & & $\kw{iffail}(L)$ \\
$\tau(A,\: L)$      &  \verb|      = | & & $\kw{call} ~A$  \\
&  & & $\kw{iffail}(L)$ \\

$\tau(e ~ e',\: L)$ &  \verb|      = | & & $\tau(e,\: L)$  \\
 &  & & $\tau(e',\: L)$  \\

$\tau(e ~ / ~ e', L)$ &  \verb|      = | & & $\kw{push}$  \\
& & & $\tau(e,\: L')$  \\
&  & & $\kw{pop}$  \\
&  & & $\kw{jump} ~L''$  \\
&  & L' & $\kw{peek}$  \\
&  &  & $\kw{pop}$  \\
&  &  & $\kw{succ}$  \\
& & & $\tau(e',\: L)$ \\
& & L'' & $\kw{nop}$  \\


$\tau(\&e, L)$ &  \verb|      = | & & $\kw{push}$  \\
& & & $\tau(e,\: L')$  \\
&  & L' & $\kw{peek}$  \\
&  &  & $\kw{pop}$  \\
&  &  & $\kw{iffail} ~L$  \\

$\tau(!e, L)$ &  \verb|      = | & & $\kw{push}()$  \\
& & & $\tau(e,\: L')$  \\
&  &  & $\kw{peek}$  \\
&  &  & $\kw{pop}$  \\
&  &  & $\kw{fail}$  \\
&  & & $\kw{jump} ~L$  \\
&  & L' & $\kw{peek}$  \\
&  &  & $\kw{pop}$  \\
&  &  & $\kw{succ}$  \\ \hline

\end{tabular}
\caption{Definition of the conversion function $\tau(e)$}
\label{fig:convert}
\end{center}
\end{figure}

As described in Section 3.2, our parsing instructions are based on the static jump to alternative. The failure-case jump to $L$ is invoked by the execution of \kw{iffail}(L). For example, \kw{iffail}(L) is executed after \kw{char}(c) to check the failure of matching. The failure status is preserved until the explicitly controlled by \kw{succ}(). This is why we need to recheck the failure status in cases of nonterminal calls. The jump analysis is closed within a production local scope.

The choice expression $e ~ / ~ e'$ is a typical case of involving backtracking handling. Likewise, the option $e?$ and $e*$ encode the similar backtracking handling. Note that these operators by nature are a syntax sugar of the choice expressions in PEGs. The $\&$ and $!$ predicates always consume no character. This means we need to rollback the parsing position in cases of either success or failure. In the not-predicate, we turn over the result status. 

\section{MiniNez}

MiniNez is a PEG virtual machine for processing text on embedded systems. In this section we describe the implementation of MiniNez. 

\subsection{Architectural Overview}

The machine virtualization approach requires the pre-compilation to parsing instructions. As we described in Section 3.3, converting parsing expressions into parsing instructions is fairly simple. However, converted code is neither the best nor good enough in reality. Advanced code refinement, or optimization, is required both to improve the performance and to reduce code size. The integrated compilation on the machine might be desirable in terms of usability, but perhaps too costly in embedded system environments.

Instead, we have adopted the decoupled architecture of virtual machine from bytecode compiler. As in Java virtual machine and other embedded virtual machine implementations\cite{TinyKonoha}, we can expect a smaller footprint of implemented virtual machine. As a result, our implemented MiniNez tool consists of the following two parts:

\begin{itemize}
\item MiniNez compiler
\item MiniNez machine
\end{itemize}

\begin{figure}[t]
	\begin{center}
		\includegraphics[width=60mm]{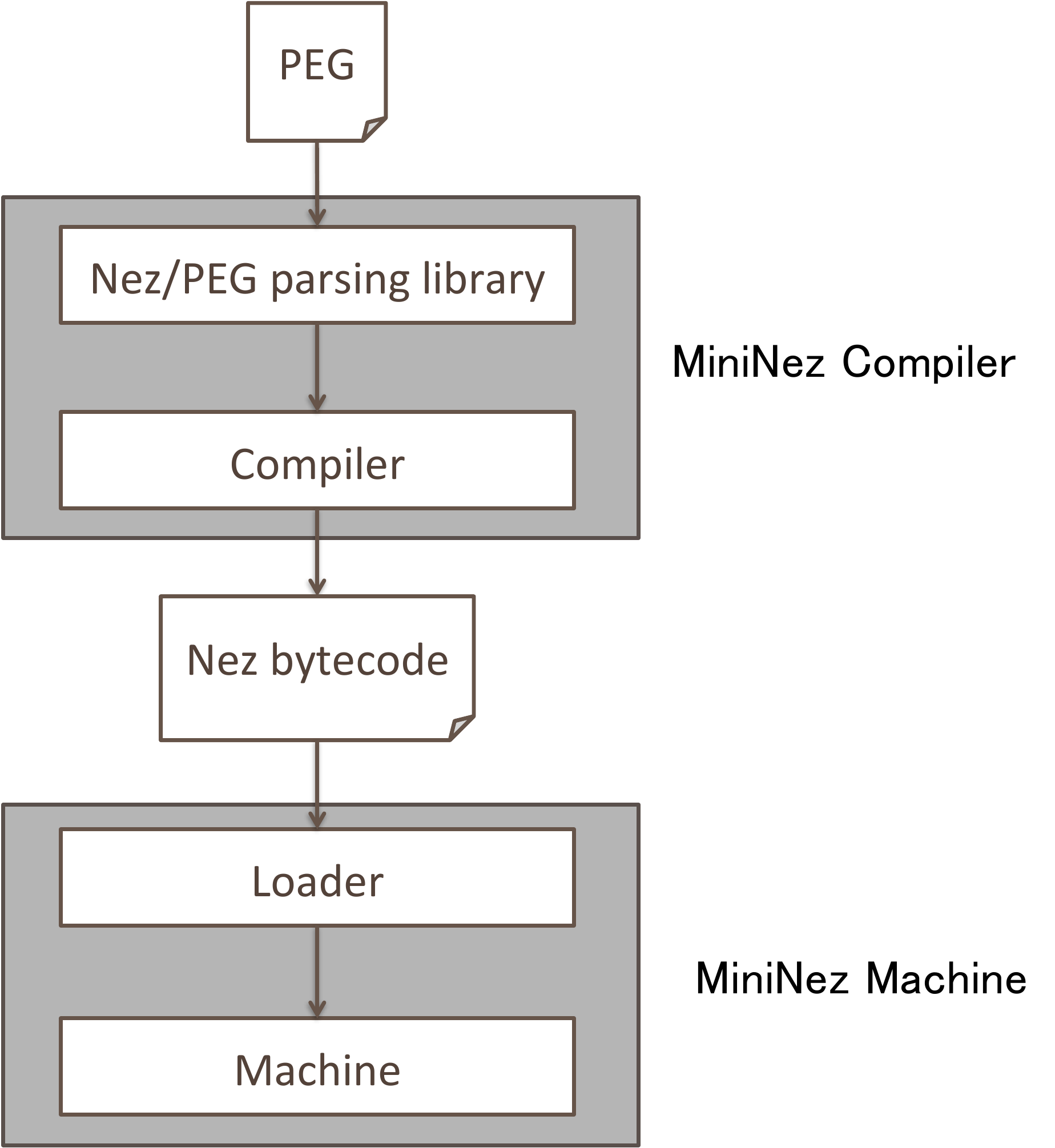}
		\caption{MiniNez architectural overview}
		\label{fig:nezvm_composition}
	\end{center}
\end{figure}

Figure \ref{fig:nezvm_composition} shows the architectural overview of MiniNez. The users use a PEG language to specify their grammars. The MiniNez compiler loads the specified grammar file and produces a compiled code file. The MiniNez machine loads the compiled code and instantiates a parser runtime when parsing the input string. Unlike parser generators, any C code generation and C compilation are necessary for loading new grammars. 

MiniNez compiler is written in Java and based on our developed Nez parsing tool kit, including a PEG parser, a grammar analysis framework, and code generator framework. That is, we generate some optimized MiniNez code, instead of C code generation for the specified parser. In Section 5, we perform the comparative study with C parsers that are generated by Nez parsing toolkit\cite{ASTMachine}. MiniNez machine is written in C and in a very portable manner without any library dependencies. Since MiniNez code is a platform-independent, MiniNez can run in many resource-restricted computing environments, including Linux Kernel. 

\subsection{MiniNez Compiler}

MiniNez compiler integrates optimization opportunities to refine converted code. In this section, we focus  mainly on downsizing optimizations, including {\em instruction specialization}, {\em cost-based inlining}, and {\em static flow analysis}.

\subsubsection{Instruction Specialization}

Instruction specialization is an optimization technique that replaces frequently appearing instruction sequences, or {\em sequence patterns}, with a single specialized instruction. Apparently, this optimization allows the size reduction of compiled code. 
Figure \ref{fig:specialization_sample} shows an example of instruction specialization in case of \verb|!c|. The left-hand side is a code fragment converted by $\tau(!c, L)$, and the right-hand side is the same code replaced with a specialized instruction \kw{nchar}(c), standing for not-character matching. In this example, 10 instructions are reduced into a single instruction. 

\begin{figure}[t]
\begin{center}
\begin{tabular}{ll|ll} 
 & unspecialized & & specialized \\ \hline
 & $\kw{push}()$    & & $\kw{nchar}('c', L)$ \\
 & $\kw{char}('c')$   & & \\
 & $\kw{iffail}(L')$    & & \\
 & $\kw{peek}()$     & & \\
 & $\kw{fail}()$        & & \\
 & $\kw{jump}(L)$   & & \\
 L' & $\kw{peek}()$ & & \\
  & $\kw{pop}()$     & & \\
  & $\kw{succ}()$    & & \\
 & $\kw{nop}()$      & & \\ \hline
\end{tabular}
\end{center}
\caption{Example of instruction specialization}
\label{fig:specialization_sample}
\end{figure}

As shown above, the instruction specialization has a huge effect on code reduction. On the other hand, an indiscriminate increase of specialized instructions simply fatten the virtual machine footprint. We have carefully designed by three approaches: peephole specialization, lexical specialization, and unary specialization.

Peephole specialization is based on the frequency analysis of paired instructions. To begin with, we pose the (at-most) {\em one-argument} constraint in designing specialized instructions. The reason is that the constraint makes it easier for us to minimize a fixed-length instruction layout, as we will describe in the next subsection. That is, for example, a sequence \kw{char}(a) \kw{iffail}(L), encoded from a character terminal {\tt 'a'}, can be {\em not} specialized due to two arguments. On the other hand, a sequence \kw{peek}() \kw{pop}() can be specialized into a single \kw{peekpop}() instruction. 

Lexical specialization is an enhancement of \kw{char}(c) instruction, which improves lexical matching capability. The instruction design is based on the analysis of the terminal equality, as follows:

\begin{itemize}
\item \kw{str}(s) -- string matching, specialized from the sequence of characters {\tt 'a'} {\tt 'b'} {\tt 'c'}
\item \kw{cmap}(m) - character bitmap-based matching, specialized from the choice of characters {\tt 'a'} / {\tt 'b'} / {\tt 'c'}
\end{itemize}

Unary specialization is based on the observation that unary expressions are likely converted with many instructions. To start, we focus on unary expressions including lexical patterns such as \kw{char}, \kw{str}, and \kw{cmap}. Table\ref{table:unary} shows the selected combinations in each unary expression. The instruction in the gray cell stands for unimplemented due to fewer appearance in real grammars.

\begin{table}[t]
	\caption{The selection of unary specialization}
	\label{table:unary}
	{\centering
		\begin{tabular}{|l|l|l|l|} \hline
			unary & \kw{char} & \kw{str} & \kw{cmap} \\ \hline
			not & \kw{nchar} & \kw{nstr} & \cellcolor[gray]{0.8} ncmap \\ \hline
			and & \cellcolor[gray]{0.8} achar & \cellcolor[gray]{0.8} astr & \cellcolor[gray]{0.8} acmap \\ \hline
			option & \cellcolor[gray]{0.8} ochar & \kw{ostr} & \kw{ocmap} \\ \hline
			repetition & \cellcolor[gray]{0.8} rchar & \cellcolor[gray]{0.8} rstr & \kw{rcmap}
			\\\hline
		\end{tabular} \\
	}
\end{table}

\subsubsection{Cost-based Inlining}

The {\em inline expansion} replaces a non-terminal call with the body of associated expressions. Simply, it can remove the cost of two instructions: \kw{call} and \kw{ret}. While inlining is a known technique for a significant performance improvement\cite{PLDI06_Rats}, the indiscriminate lining enlarges the resulting code size due to code duplication. We highlight the downsizing effect in this paper, and then apply inlining on the following conditions:

\begin{itemize}
	\item Associated expressions can be compiled to less than two instructions
	\item Nonterminals are referred from a single unique context.
\end{itemize}

Due to the above conditions, the inlining process reduces compiled code size. 

\subsubsection{Static Flow Analysis}

Static flow analysis is an optimization technique based on the control flow of failure cases. Intuitively, the conversion creates redundant instructions such as duplicated push and  jump. The following is the converted code for $(e_1/e_2)*$. The left-hand is the plain conversion by $\tau((e_1/e_2)*, L)$. The duplicated \kw{push} instructions are generated before the evaluation of the first subexpression $e_1$.  However, it is unnecessary to push the same position. The right-hand is a redundant-removed version of the left-hand. The flow analysis allows us to eliminate such duplicated operations.

\begin{center}
		\begin{tabular}{cl|cl} 
			& plain code & & removed \\\hline
			L1& $\kw{push}()$ & & $\kw{push}()$  \\
			& $\kw{push}()$ & L1& $\tau(e_1,\: L2)$ \\
			&$\tau(e_1,\: L2)$ & & $\kw{jump}(L3)$ \\
			& $\kw{pop}()$ & L2 & $\kw{peek}()$\\
			&$\kw{jump}(L3)$ & & $\kw{succ}()$  \\
			L2 & $\kw{peek}()$ & & $\tau(e_2,\: L4)$ \\
			& $\kw{pop}()$ & L3& $\kw{pop}()$ \\
			& $\kw{succ}()$ & & $\kw{jump}(L1)$ \\
			& $\tau(e_2,\: L4)$ & L4 & $\kw{peek}()$ \\
			L3 & $\kw{pop}()$ &   & $\kw{pop}()$  \\
			& $\kw{jump}(L1)$ & & $\kw{succ}()$ \\
			L4 & $\kw{peek}()$& & $\kw{nop}()$ \\
			 & $\kw{pop}()$  \\
			& $\kw{succ}()$  \\
			& $\kw{nop}()$  \\
		\end{tabular} \\
\end{center}

\subsection{MiniNez Machine}

MiniNez machine is a small virtual machine that specifically performs parsing instructions. The major difference from application virtual machines such as JavaVMs is that MiniNez provides no instruction supports for general-purpose operations. Table \ref{table:full_instrucsions} lists the full instructions of MiniNez. Currently, only 20 instructions including extended instructions in Section 4.2 are implemented on MiniNez. 

\begin{figure}[t]
	\begin{center}
		\includegraphics[width=60mm]{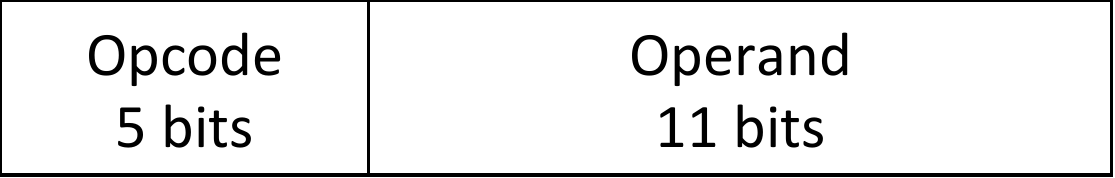}
		\caption{structure of MiniNez instruction}
		\label{fig:nezvm_instruction}
	\end{center}
\end{figure}

The small set of the supported instructions enables us to reduce the opcode layout down to 5 bits. In addition, we make the fixed length of all instructions 2-byte in length by carefully designing instruction arguments. Pointers to strings and long jump points are all reduced to 11-bit length. Figure \ref{fig:nezvm_instruction} shows the layout of instructions.

\begin{table}[t]
	\caption{Full instructions}
	\label{table:full_instrucsions}
	\begin{center}
		\begin{tabular}{lll}
			Opcode & Argument & Description \\ \hline
			\kw{nop} & &
			Do nothing \\
			\kw{succ} & &
			Succed always \\
			\kw{fail} & &
			Fail always \\
			\kw{char} & c &
			Match the character c \\
			\kw{any} & &
			Match the any character\\
			\kw{jump} & L &
			Branch always \\
			\kw{iffail} & L &
			Branch if the result is {\tt F} \\
			\kw{call} & L &
			Call the nonterminal call \\
			\kw{ret} &  &
			Return the nonterminal call \\
			\kw{push} &  &
			Push $pos$ to the stack \\
			\kw{pop} &  &
			Pop the top operand stack value \\
			\kw{peek} &  &
			Store the top operand stack value \\ 
			\kw{str} & s &
			Match the string  \\
			\kw{cmap} & m &
			Match the character bitmap map  \\
			\kw{nchar} & c &
			Not-character matching\\
			\kw{nstr} & s  &
			Not-string matching\\
			\kw{ostr} & s &
			Optional string matching\\
			\kw{ocmap} & m &
			Optional character map\\
			\kw{rcmap} & m &
			Repetition of character map\\
			\kw{peekpop} & &
			\kw{peek} and \kw{pop}\\
			\hline
		\end{tabular} \\
		s: string, m: character bitmap
		\end{center}
\end{table}

The body of the MinNez machine is implemented with a single C function that iterates instruction executions. Figure \ref{fig:function} illustrates our implemented code that is excerpted from the virtual machine function. All paring states are bounded as local variables, ({\tt pc}, {\tt pos}, {\tt sp}, {\tt result}), which can be statically allocated to registers by modern C compilers. The dispatch method of instructions that we have adopted is the {\em indirect threaded coding}\cite{PLDI98_VM}, a table-based mapping from an opcode to the jump address. In total, the size of a virtual machine is approximately 100 lines in C source code, which contains no function calls to external libraries. 

\begin{figure}[tb]
{\small \begin{framed} 
\begin{verbatim}
  int run(Instruction *code , Context *ctx) {
    Instruction *pc;
    long pos;
    long *sp;
    int r;   /* result */
    void *jump_table[] 
      = {&&I_char, &&I_iffail ... };
    I_char:
      if (pc->arg != *ctx->inputs[pos++]) {
        --pos;
        r = 1;
      }
      goto *(jump_table[(pc++)->op]);
    I_iffail:
      if (!r) {
        goto *(jump_table[(pc+=pc->arg)->op]);
      }
      goto *(jump_table[(pc++)->op]);  

    ...  
      
    return r;
  }
\end{verbatim} \end{framed}}
\caption{The C Virtual Machine Function (Excerpted)}
\label{fig:function}
\end{figure}

The first argument {\tt Instruction} is a reference to a loaded bytecode, and the second argument {\tt Context} is a reference to a runtime information that is required to parse an input. The parsing context includes an input string and a VM stack to be operated by parsing instructions. The stack is allocated with a given size by the user at each parser instantiation time. The default size is set to 2K, which corresponds to more than 500 recursive nonterminal calls. 

\section{Experimental Results}

This section describes the memory consumption and parsing performance of MiniNez. 

\subsection{Experimental Setup}

We assume that main parser applications on embedded systems are data parsing and input validations. To obtain practical insights on such applications, we prepare 6 distinct grammars (CSV, LOG, XML, JSON, Email, and UTF8), where the former four grammars are intended for data parsing and the latter two are for input validations. Data sets that we set up are collected:

\begin{itemize}
\item CSV -- open data distributed from usgs.gov. 
\item LOG -- syslog files randomly collected from MacOS X.
\item XML --  synthetic and scalable XML files that are provided by XMark\cite{VLDB02_XMark} benchmark project.
\item JSON -- open data formatted in GeoJSON, distributed from usgs.gov.
\item Email -- EDRM Enron email data sets
\item UTF8 -- Japanese Wikipedia articles that is provided by Wikipedia.
\end{itemize}

It is important to note that PEGs's backtracking causes an potential exponential parsing time in worst cases. However, it is a known fact that such exponential cases rarely happen in practice. We have selected grammars that involves no exponential parsing behaviors as reported in \cite{PRO101}.

MiniNez is written in a portable C and compiled by GCC4.4 on a Linux operating system. To compare, we set up two C generated parsers by Leg\cite{Leg} and our Nez-C parsing tool\cite{PRO101}. These generated parsers are implemented with a plain recursive decent parsing. Note that C generated parser were executed with a standard stack size of Linux operating system (8MB) in order not to consume C stack by recursion. 

Tested computing environments cover the following popular architectures:

\begin{itemize}
	\item Raspberry Pi 2 Model B, with 900MHz quad-core ARM Cortex-A7, 1GB RAM, running on Linux
	\item Edison, with 500MHz dual-core Intel Atom, 1GB RAM, running on Linux
\end{itemize}

\subsection{Memory Consumption}

To begin with, we examine the memory consumption of MiniNez. Table \ref{table:opt_inst_size} shows the size of compiled code in each grammar. The column labeled "Rules" indicates for the size of production rules in the grammar. The column labeled "Plain" indicates no optimized code, while the column "Reduced" indicates fully deployed code reduction described in Section 4.2. 

\begin{table}[tbh]
	\caption{Size of compiled code}
	\label{table:opt_inst_size}
	{\centering
		\begin{tabular}{cccc}
			Grammar & Rules & Plain code[byte] & Reduced code[byte]   \\ \hline
			CSV & 4 & 198 & 84\\
			LOG & 11 & 1458 & 350\\
			JSON & 16 & 2288 & 368 \\
			XML & 21 & 2048 & 366 \\
			Email & 5 & 1146 & 66 \\
			UTF8 & 2 & 1626 & 38 \\
			\hline
		\end{tabular} \\
	}
\end{table}

\begin{figure}[tbh]
	\begin{center}
		\includegraphics[width=70mm]{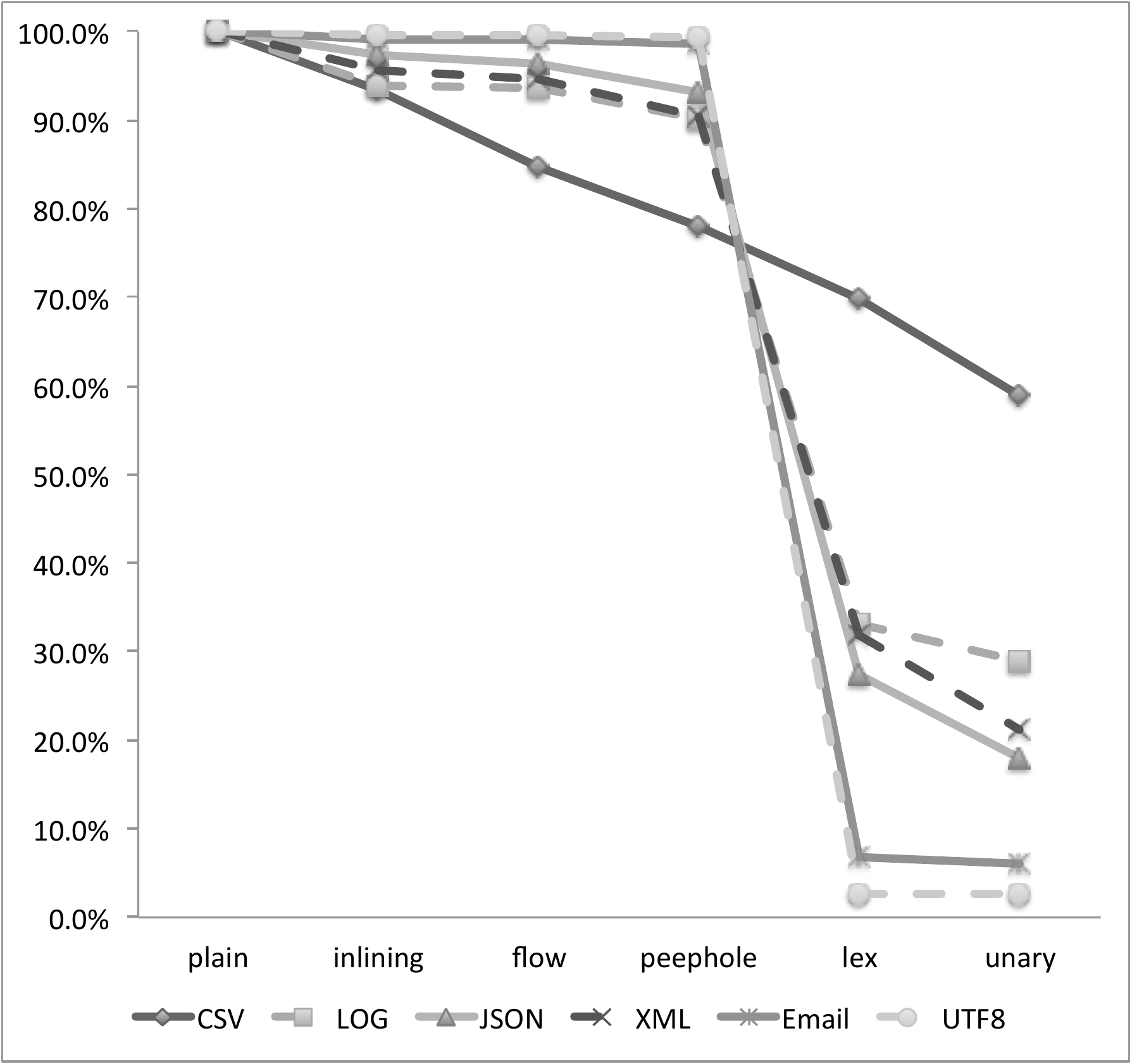}
		\caption{Downsizing effects of each optimization}
		\label{fig:optimize_effect}
	\end{center}
\end{figure}

\begin{table}[tbh]
\begin{center}
	\caption{Stack consumption of MiniNez}
	\label{table:stack}
	{\center
		\begin{tabular}{cr|r}
			Grammar & File size & Used stack [byte] \\\hline
			CSV & 1.3MB & 24 \\
			LOG & 3.1MB & 28 \\
			JSON & 6.1MB &128 \\
			XML & 5.0MB & 208 \\
			Email & 144MB & 340 \\
			UTF8 & 82MB & 12 \\\hline
		\end{tabular} \\
	}
\end{center}
\end{table}

Figure \ref{fig:optimize_effect} shows the effects of downsizing by comparing the size of compiled code in proportion to the "plain" results. The data point labeled "non-opt" indicates no optimization, while other data points indicates each optimizations. The data points labeled {\tt inline}, {\tt flow}, {\tt peephole}, {\tt lex}, and {\tt unary}, respectively, indicate the deployment of cost-based inlining (in Section 4.2.2), static flow analysis (in Section 4.2.3), peephole specialization, lexical specialization and unary specialization (in Section 4.1). The graphed optimization effects are cumulative from left to light. That is, the data point labeled {\tt flow} includes {\tt inlining} and {\tt flow}. We confirm that lexical and unary specialization are two best effective for code reduction.

We will turn to the runtime memory consumption of MiniNez. Since MiniNez is based on a recursive decent parsing, the stack consumption in nonterminal calls is a major factor. Table \ref{table:stack} shows the VM stack consumption of MiniNez. The column labeled "file" indicates the input size of tested file, and the "used stack" indicates the maximum consumption of VM stack during parsing. We confirm that the VM stack of at most 320 bytes can parse even the large files. Note that the necessary stack size depends on grammars, not the size of inputs. 


\begin{table}[tbh]
\begin{center}
	\caption{Binary size of MiniNez, Leg, and Nez-C}
	\label{table:comparison_leg}
		\begin{tabular}{c|cccc}
			Environment & Grammar & MiniNez [KB] & Leg[KB] & Nez-C[KB]   \\ \hline
			& CSV & 9.0 & 18.7 & 21.4\\
			& LOG & 9.3 & 24.0 & 28.3 \\
			Raspberry Pi 2 & JSON & 9.3 & 32.0 & 34.6\\
			& XML & 9.3 & 32.6 & 37.5 \\
			& Email & 9.0 & 24.6 & 30.5 \\
			& UTF8 & 8.9 & 18.3 & 24.6 \\\hline
			& CSV & 10.0 & 24.6 & 25.1\\
			& LOG & 10.3 & 33.9 & 34.2 \\
			Edison & JSON & 10.3 & 37.6 & 35.4 \\
			& XML & 10.3 & 40.9 & 36.7\\
			& Email & 10.0 & 29.3 & 31.2 \\
			& UTF8 & 10.0 & 22.4 & 26.5 \\
			\hline
		\end{tabular} \\
\end{center}
\end{table}

Table \ref{table:comparison_leg} shows a comparison of the binary size of MiniNez and C parsers that is generated by Leg and Nez. MiniNez includes the size of bytecode to compare each grammar. We confirm that MiniNez is approximately 20KB smaller than each of generated parsers. Since MiniNez can switch syntax by only loading bytecode, MiniNez achieves a very small footprint in parsing tasks. 

\subsection{Performance Evaluation}


\begin{table}[tbh]
\begin{center}
	\caption{Throughputs in byte per sec [MiB/s] }
	\label{table:performance}
		\begin{tabular}{c|c|cccccc}
			  Environment & Parser & CSV & LOG & JSON & XML & Email & UTF8 \\\hline
			  & MiniNez  & 6.18 & 4.82 & 2.99 & 5.88 & 1.68 & 6.11 \\
			  Raspberry Pi 2 & Nez & 5.28 & 3.45 & 2.73 & 4.02 & 1.34 & 5.88 \\
			  & leg & 3.90 & 3.70 & 2.86 & 3.46 & 1.55 & 3.74\\\hline
			  & MiniNez & 6.07 & 4.11 & 2.42 & 5.71 & 0.95 & 3.81 \\
			  Edison & Nez & 5.14 & 3.72 & 2.02 & 3.68 & 0.81 & 3.26\\
			  & leg & 3.71 & 3.60 & 2.37 & 3.39 & 0.89 & 3.43\\
		\end{tabular}
\end{center}
\end{table}

Finally, we will examine the parsing performance. Table \ref{table:performance} shows the CPU time required for each type of analysis, giving the average times measured across 5 runs. We also show the throughput in mega-bytes/sec. MiniNez has throughput 4.2[MB/s] on the average, while Leg parsers have 3.1[MB/s] and Nez-C parsers have 3.4[MB/s] against the same inputs. This result suggests that MiniNez has practical and competitive performance, compared to generated parser.

\section{Related Work}

Parsing and syntactic analysis is ubiquitous and an integral part of text processing. Since the development of Lex/Yacc, the parser generator has increasingly adopted as a standard approach to formal grammar engineering. In contexts of data parsing, binpac\cite{IMC06_Binpac} and XMLScreamer\cite{WWW06_XMLScreamer} notably extend Yacc to produce efficient protocol parsers and XML parsers, respectively. However, Yacc and its extended tools do not guarantee its availability on resource-restricted embedded systems, while they produce a portable parser code. To our knowledge, the parser generation that is specialized for embedded systems has been unexplored, and then is an important open problem to be addressed. 

PEGs\cite{POPL04_PEG} are a relatively newer formalism than Context Free Grammars, and have received much popularity among many parser researchers. There are many PEG-based parser tools such as Pappy\cite{ICFP02_PackratParsing}, Rats\!\cite{PLDI06_Rats}, Mouse\cite{FI07_Mouse}, Leg/Peg\cite{Leg}, PEGjs\cite{PEGjs}, LPeg\cite{LPeg}, Waxeye\cite{Waxeye} and so on.  Most of these tools are based on a parser generator as with in Lex/Yacc. An important exception is LPeg, a pattern matching library for Lua\cite{LPeg,Lua}. As with MiniNez, LPeg has adopted a virtual parsing machine \cite{DLS08_LPeg} instead of parser generations.  MiniNez and LPeg significantly differ in terms of backtracking handling: local static jump vs. global dynamic jump. Since the failure jump address is not pushed on the stack, MiniNez achieves a fewer stack consumption. In addition, we add several specialized downsizing to the MiniNez design and implementation. LPeg, on the other hand, does not work standalone and requires a complementary Lua runtime, resulting in hard comparison on the same condition. 

\section{Conclusion}

PEGs are a formal grammar foundation for describing syntax, and are not hard to generate parsers with a plain recursive decent parsing. However, the large amount of C-stack consumption in the recursive parsing is not acceptable especially in resource-restricted embedded systems. Alternatively, we have attempted the machine virtualization approach to PEG-based parsing. MiniNez, our implemented virtual machine, is presented in this paper with several downsizing techniques, including instruction specialization, inline expansion and static flow analysis. As a result, the MiniNez machine achieves both a very small footprint and competitive performance to generated C parsers. We have demonstrated the experimental results by comparing on two major embedded platforms: Cortex-A7 and Intel Atom processor.

The applications of PEGs on embedded systems would provide several opportunities for protocol filtering and schematic data validations. Future work that we will investigate includes a declarative extraction of structured data from parsed results, enabling an alternative standard tool to regular expressions. Linear parsing guarantee is not visited in this paper, an efficient packrat parsing in restricted-memory environments is still an open challenge. The prototype implementation of the MiniNez is available online at \url{http://github.com/nez-peg/nezvm1}.

\section*{Acknowledgment}
The authors thank the IPSJ/SIGPRO members for their feedback and discussions. 

\bibliography{../bib/parser,../bib/mypaper,../bib/data,../bib/url}  

\end{document}